\documentclass{ws-procs961x669}            
\usepackage{ws-toc,ws-multind}
\usepackage{xspace}
\usepackage{hyperref}
\usepackage{journalnames}

\newcommand{\pdx}[1]{_{\rm #1}}
\newcommand{\rev}[1]{\textcolor{black}{#1}}

\def\pdot {\dot P}

\def\edot {\dot E}

\def\nh{$N_{\rm H}$\xspace}

\def\col{cm$^{-2}$\xspace}

\def\sps{s\,s$^{-1}$\xspace}

\def\flux{erg\,s$^{-1}$\,cm$^{-2}$\xspace}
\def\lum{erg\,s$^{-1}$\xspace}

\def\deg{^\circ}

\def\xds{XDINSs\xspace}
\def\xd{XDINS\xspace}

\newcommand{\psrj}{PSR J0726$-$2612\xspace}
\newcommand{\rxj}{RX J1856.5$-$3754\xspace}

\def\xmm{{\em XMM-Newton}\xspace}

\def\chandra{{\em Chandra}\xspace}
\def\rosat{{\em ROSAT}\xspace}
\def\erosita{{\em eROSITA}\xspace}
\def\nicer{{\em NICER}\xspace}

\usepackage{color}

\makeindex{author}                     

\begin{document}








\wstoc{X-ray emission from isolated neutron stars: latest results from XMM-Newton, NICER and eROSITA}{M. Rigoselli}

\title{X-ray emission from isolated neutron stars: latest results from XMM-Newton, NICER and eROSITA}

\author{M. Rigoselli}

\address{INAF OA-Brera,
via Brera, 28\\
Milano 20121, Italy\\
and\\
INAF IASF-Milano,
via A. Corti, 15\\
Milano 20133, Italy\\
\email{michela.rigoselli@inaf.it}}

\begin{abstract}
The X-ray spectra of isolated neutron stars (INSs) typically include a thermal component, that comes from the cooling surface, and a non-thermal component, produced by highly-relativistic particles accelerated in the stellar magnetosphere. Hot spots from returning currents can also be detected.

Middle-aged pulsars exhibit a mixture of these components, but other flavours of INSs, that show a large variety of physical parameters (such as spin period, magnetic field and age) emit only thermal X-rays. Typically, these stars are detected either in large serendipitous datasets from pointed X-ray observations or from searches in the data of all-sky surveys.

The connection between these thermally-emitting INSs, the ordinary pulsars, and the new emergent class of pulsars characterized by a long period, that do not show X-ray emission despite their high magnetic field, is one of the current challenges in the study of neutron stars.

In this contribution I will review the latest results on several objects belonging to various INS classes, such as the XDINS \rxj, the enigmatic Calvera, the long period PSR J0250+5854 and the new thermal INS candidates, obtained with the X-ray observatories \xmm, \nicer and \erosita.
\end{abstract}

\bodymatter

\section{The Isolated Neutron Stars zoology}
Neutron stars were first discovered as radio pulsating sources in 1967 \citep{hew68}, and in the following half a century more than 3700 have been recorded\footnote{\url{https://www.atnf.csiro.au/research/pulsar/psrcat/}.} \citep{2005AJ....129.1993M}. They have been discovered mainly by the detection of their pulsed non-thermal emission, at wavelengths ranging from radio to $\gamma$-rays, and may be isolated stars or members of a binary system.

The energy that sustains pulsar emission is supplied by their fast rotation, via the braking operated by their intense magnetic field. At the beginning of their life, pulsars spin with periods $P_0$ of the order of milliseconds, and they gradually slow down. Under the assumption of a constant, dipolar magnetic field and that the current period $P$ is larger than $P_0$, three characteristic quantities can be inferred from $P$ and its derivative $\pdot$: the so-called spin-down luminosity
\begin{equation}
    \dot{E}_{\rm rot} = 4 \pi^2 I_{\rm NS} \dot{P} P^{-3},
    \label{eq:erot}
\end{equation}
where $I_{\rm NS} \approx 10^{45}$ g cm$^2$ is the moment of inertia of the neutron star, the dipolar magnetic field on the surface
\begin{equation}
    B_{\rm dip} \approx 3.2 \times 10^{19}\, (P \dot{P})^{1/2}~\rm G,
    \label{eq:bdip}
\end{equation}
and the characteristic age
\begin{equation}
    \tau_c = \frac{P}{2\dot{P}}.
    \label{eq:tauc}
\end{equation}

\begin{figure*}[!ht]
\centering
\includegraphics[width=1\textwidth]{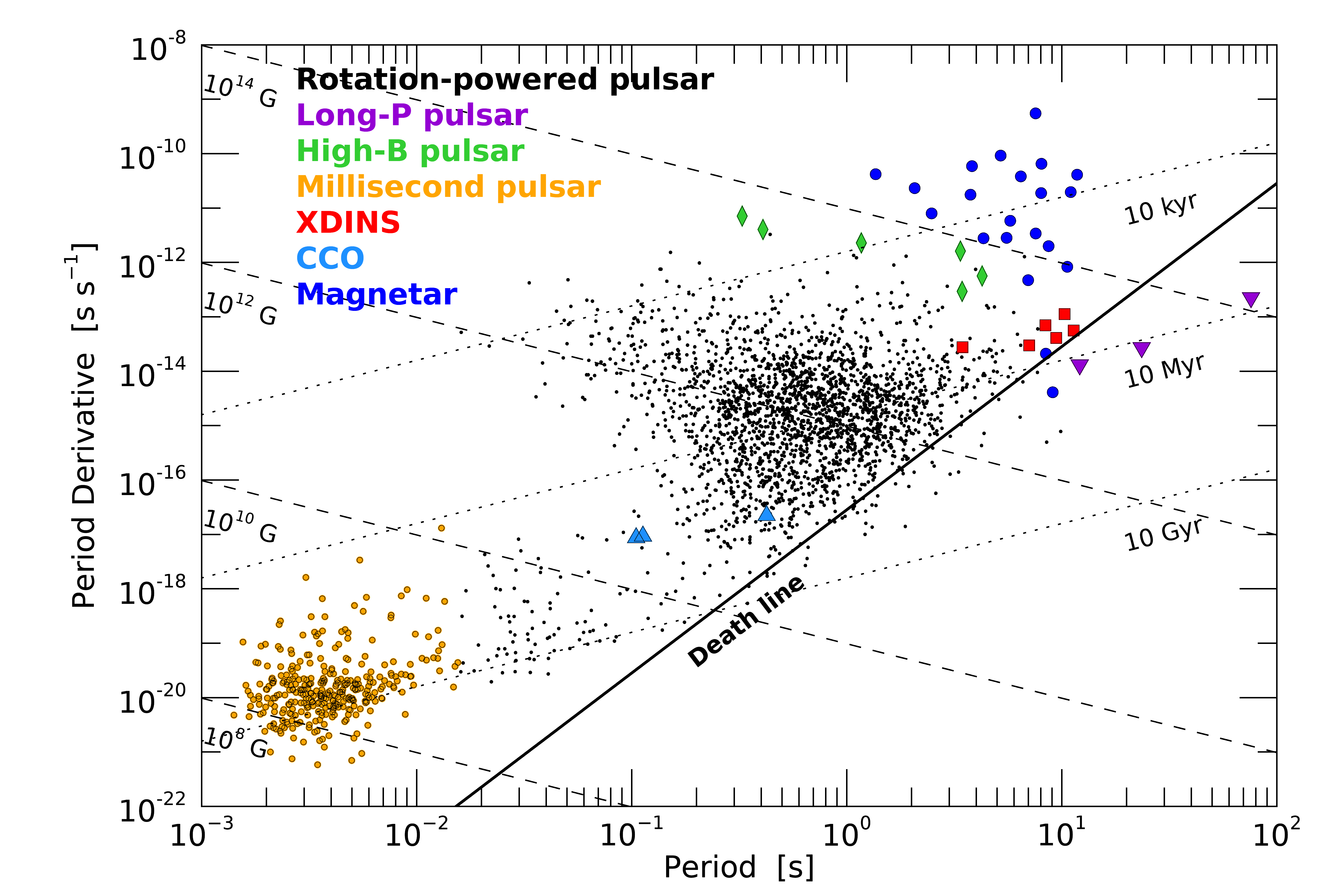}
\caption{\small $P-\pdot$ diagram of pulsars. The bulk of the pulsar population is made up of the rotation-powered pulsars (black dots), sub-divided into long-P pulsars (purple triangles), millisecond pulsars (orange dots), and high-B pulsars (green diamonds). The other classes of INSs plotted here are the magnetars (blue dots), the \xds (red squares) and the CCOs (light blue triangles). Lines of equal characteristic age (dotted, $10^{4} - 10^{10}$ yr) and equal dipole magnetic field (dashed, $10^{8} - 10^{14}$ G) are indicated. The radio pulsar death line $B/P^2 = 1.7 \times 10^{11}$ G s$^{-2}$ of Bhattacharya et al.\ 1992 \citep{bha92} is also shown. The data are taken from the ATNF Pulsar Catalogue, version 2.5.1 \citep{2005AJ....129.1993M}.} 
\label{fig:PPdot}
\end{figure*}

$P$ and $\pdot$ play a fundamental role in characterizing the pulsar properties, and the neutron star population is usually represented in the $P-\pdot$ diagram (shown in Figure~\ref{fig:PPdot}), as the ordinary stars are represented in the Hertzsprung-Russell diagram; the different classes of neutron stars are placed on different zones of the diagram.

In the following, I will review the main X-ray properties of isolated neutron stars (INSs) \citep{2013FrPhy...8..679H,2014AN....335..262I,2018IAUS..337....3K,2023Univ....9..273P} that are powered by rotation (Section~\ref{sec:rpp}) or by secular cooling (Section~\ref{sec:cooling}), and recent discoveries that assess the links between the different classes (Sections~\ref{sec:new} and~\ref{sec:multi}).

\subsection{Rotation-powered pulsars}\label{sec:rpp}

Rotation-powered pulsars (RPPs) are the bulk of INSs and are detected thanks to their pulsed emission.
Nowadays more than 3500 RPPs are known, and they have been detected from the radio band to the very-high energy $\gamma$-rays (see e.g.\ the recent result on Vela pulsating emission up to 20 TeV \citep{2023NatAs.tmp..208H}). 
RPPs fill the central region of the $P-\pdot$ diagram (Figure~\ref{fig:PPdot}, black dots); a newborn pulsar appears in the top-left corner and, according to the magnetic-dipole braking model, it evolves along the $B\pdx{dip}$ constant lines (dashed), crossing the $\tau_c$ constant lines (dotted). The bottom-right corner of the plot is empty because when a pulsar crosses the so-called ``death line'' \citep{1993ApJ...402..264C}, it is too old and too slow to sustain the required potential drop for pair production in the vacuum gap \citep{bar98,bar01}.

Actually, a growing number of pulsars with $P>12$ s that crosses the death line (see Figure~\ref{fig:PPdot}, purple triangles) has been discovered in the last few years (see Section~\ref{sec:longp}). It should be noted that there are many possible death lines (see the discussion in Suvorov \& Melatos 2023\citep{2023MNRAS.520.1590S}), which differ for the emission model (inner vacuum-gap curvature radiation or space-charge-limited flow), the nature of the seed $\gamma$-ray photons for pair production (curvature radiation or Inverse Compton), the magnetic field configuration (dipolar, multipolar, twisted) and the intensity of the frame dragging. Nevertheless, the observed timing and polarization properties, as well as the radio and X-ray manifestations, challenge the current picture of INSs.

RPPs can be further divided into old, fast spinning pulsars (the millisecond pulsars \citep{man17}), and young, strongly magnetized pulsars (the high-B pulsars \citep{ng11}). 
The millisecond pulsars are old pulsars that nonetheless have short and stable periods \rev{($P \lesssim 10$ ms}, $\pdot \lesssim 10^{-18}$ s s$^{-1}$, see Figure~\ref{fig:PPdot}, orange dots) that have been spun up through the accretion of matter from the binary companion \rev{over a prolonged phase lasting even giga-years in their evolutionary history}. The majority of them still have a companion star and emit powerful non-thermal emission in radio, X- and $\gamma$-ray bands, \rev{and sometimes an additional thermal X-ray component from heated polar caps.}

The high-B pulsars have magnetic fields higher than the quantum critical field
\begin{equation}
    B\pdx{QED} = \frac{m_e^2c^3}{e \hbar} \approx 4.4\times10^{13} \mathrm{~G}
    \label{eq:bqed}
\end{equation}
(see Figure~\ref{fig:PPdot}, green diamonds), and they do not constitute a separated class of INSs, but they act as ordinary pulsars most of the time, and suddenly show bursting phenomena as the magnetars or share the spectral properties of the XDINSs (see Section~\ref{sec:link}).

\subsection{Cooling neutron stars}\label{sec:cooling}
Neutron stars also shine in the optical/UV/X-ray frequency range thanks to their hot surface. Thermal emission can be detected from the cooling surface and from hot spots heated by returning currents (see Potekhin et al.\ 2020 \citep{pot20} and references therein).
Typically, this component is less pulsed and less beamed than the non-thermal one, which increases the likelihood of serendipitous discoveries. Recent discoveries of new thermally-emitting INS candidates are reported in Section~\ref{sec:tins}. 

Among thermally-emitting INSs, there are two small groups that do not show magnetospheric activity  at any frequency \citep{2009ApJ...702..692K,mig19}, and it is not clear yet if they are a different flavor of INS or they are just seen with an unfavorable line of sight. They have been discovered thanks to their high X-ray-to-optical flux ratio (the X-ray-dim isolated neutron stars, \xds \citep{2007Ap&SS.308..191V,2009ASSL..357..141T}) or as hot sources at the center of supernova remnants (the central compact objects, CCOs \citep{got13,2017JPhCS.932a2006D}). Their thermal X-ray spectra can also show absorption lines at a few hundreds of eV. If these lines are interpreted as cyclotron features, 
the estimated magnetic fields are 
\begin{equation}
    B\pdx{cyc,e} = E\pdx{abs} \frac{m_e c} {\hbar e}(1 + z)^{-1} \approx  \frac{E\pdx{abs}}{100~ \mathrm{eV}} \, 7.2 \times 10^{9}~\mathrm{G}.
    \label{eq:Bcyce}
\end{equation}
in the case of electrons, or
\begin{equation}
    B\pdx{cyc,p} = E\pdx{abs} \frac{m_p c} {\hbar e}(1 + z)^{-1} \approx  \frac{E\pdx{abs}}{100~ \mathrm{eV}} \, 1.3 \times 10^{13}~\mathrm{G}.
    \label{eq:Bcycp}
\end{equation}
in the case of protons.

The \xds are seven INSs discovered in the nineties by the \rosat satellite and soon gained the nickname of ``Magnificent Seven''.
They have spin periods in the range $3-12$ s and period derivatives of a few $10^{-14}$ s\,s$^{-1}$ (see Figure~\ref{fig:PPdot}, red squares), which result in characteristic ages of $\tau_c \sim 1-4$ Myr and magnetic fields of the order of a few $10^{13}$ G.
Their very soft ($kT\lesssim 100$ eV) X-ray spectra are well reproduced by a simple blackbody with little interstellar absorption, with the additional presence of broad absorption lines at $200-400$ eV in most sources \citep{2003A&A...403L..19H, 2004ApJ...608..432V,2004A&A...424..635H,2005ApJ...627..397Z}, and narrow, phase-variable ones in few cases \citep{2015ApJ...807L..20B,2017MNRAS.468.2975B}.

The CCOs form a class that counts a dozen objects. Their thermal spectra have high temperatures ($200-500$ eV) and very small emitting radii (ranging from 0.1 to a few km), and can show absorption features at $700-800$ eV \citep{big03,got10}. Currently there are only three pulsating CCOs (see Figure~\ref{fig:PPdot}, light blue triangles), that have periods of $0.1-0.4$ s \citep{zav00,got05,got09} and spin derivatives of about $10^{-17}$ s~s$^{-1}$ \citep{hal10,got13}, from which weak dipole magnetic fields ($B\pdx{dip} \sim 10^{10}$ G) and high characteristic ages ($\tau_c \sim 10^{8}$ yr) are derived. This is at variance with the supernova remnant (SNR) associations, and the reason could be that the approximation of Eq.\ \ref{eq:tauc} is no longer valid because these sources have $P \approx P_0$. 
\rev{One remarkable CCO in SNR RCW 103 shows a 6.7-hr X-ray periodicity of yet unknown origin as well as distinctly magnetar-like behavior \citep{2016ApJ...828L..13R,2016MNRAS.463.2394D}. Such long spin periods in young non-accreting objects can be explained in a model where a strong magnetic field of the star interacts with a fallback disc (see also Section \ref{sec:longp}).}

\section{New discoveries in the X-ray band}\label{sec:new}

\subsection{Long period pulsars}\label{sec:longp}

There is a new emergent class of pulsars characterized by a long period: PSR J2251$-$3711 ($P\approx12.1$ s \citep{2020MNRAS.493.1165M}), PSR J1903$+$0433 ($P\approx14.0$ s \citep{2021RAA....21..107H}), PSR J0250$+$5854 ($P\approx23.5$ s \citep{2018ApJ...866...54T}), and PSR J0901$-$4046 ($P\approx75.9$ s \citep{2022NatAs...6..828C}). They are located near magnetars and \xds on the $P-\pdot$ diagram, therefore they are characterized by high $B\pdx{dip} \sim 10^{13-14}$ G and long $\tau_c \sim 10^7$ yr, but on the contrary they are radio sources without any X-ray counterpart. 

\begin{figure*}[ht]
\centering
\includegraphics[width=0.66\textwidth]{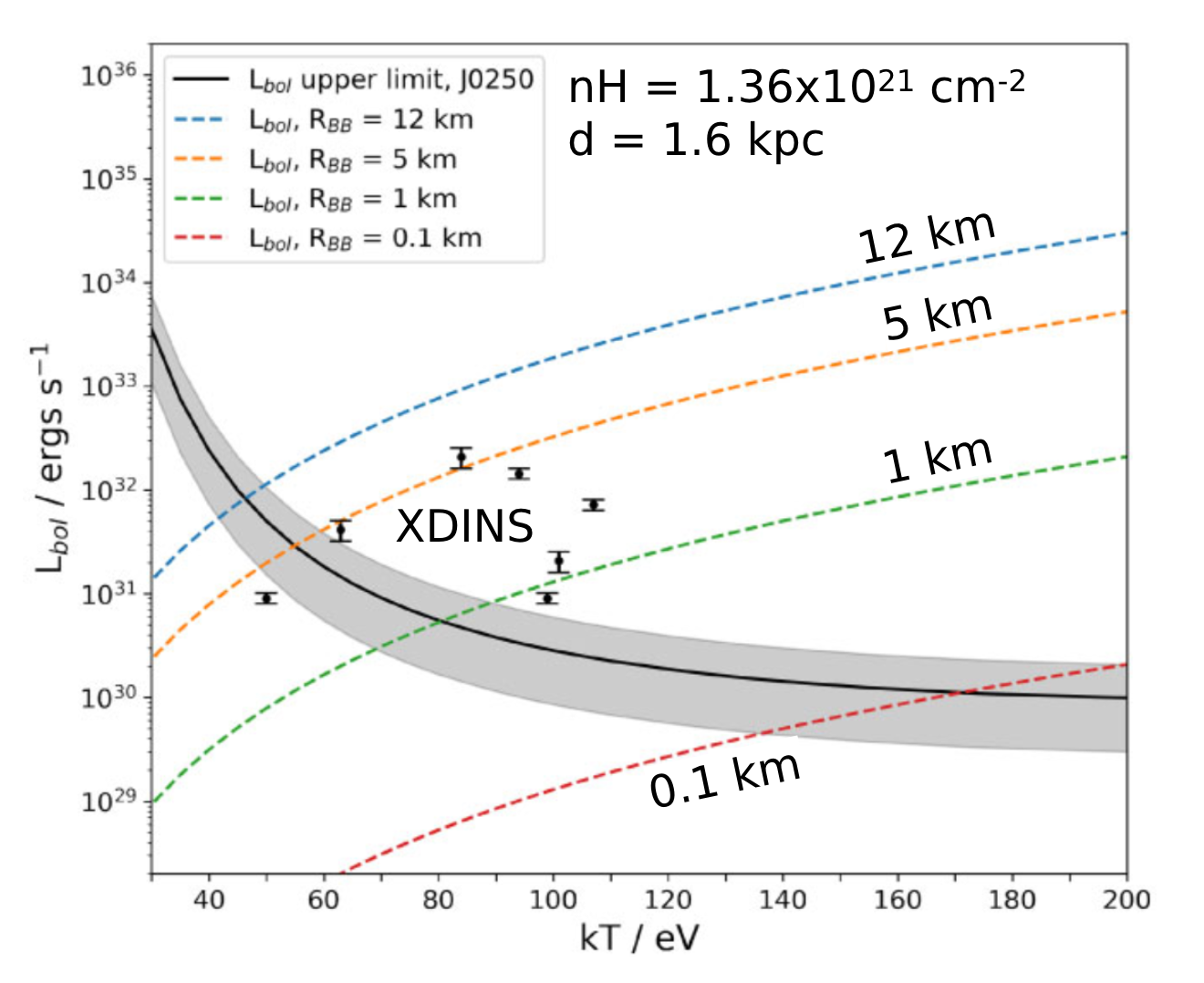}
\caption{\small Bolometric luminosity upper limit of PSR J0250$+$5854 for various values of blackbody temperature between 30 and 200 eV. The shaded region is the uncertainty in the computed upper limit, considering $d = 1.6 \pm 0.7$ kpc. The dashed colored lines are the expected bolometric luminosity of a blackbody of emitting radius 12, 5, 1, and 0.1 km, respectively. Adapted from Tan et al.\ 2023 \citep{2023MNRAS.520.5960T}.}
\label{fig:longp}
\end{figure*}

The upper limits on the X-ray luminosity confirm that they are rather old neutron stars \citep{2022ApJ...940...72R}. Tan et al.\ 2023 \citep{2023MNRAS.520.5960T} report the deep \xmm observation of PSR J0250$+$5854. In almost 100 ks of \xmm/EPIC exposure time, a $3\sigma$ upper limit \rev{on the count rate} of $9\times10^{-4}$ cts s$^{-1}$ was derived. Assuming a pure thermal emission, an absorption \nh $=1.36\times10^{21}$ \col and a distance of $1.6\pm0.7$ kpc, the bolometric luminosity upper limit shown in Figure~\ref{fig:longp} was derived. This value is below the luminosities of most of the \xds, suggesting that PSR J0250$+$5854 is unlikely a member of this class unless it has a relatively \rev{low} surface temperature ($kT<50$ eV). A similar conclusion applies if we compare PSR J0250$+$5854 with the low-B magnetars, such as SGR 0418+5729 \citep{2013ApJ...770...65R}, whose thermal luminosity comes from hot spots maintained by the bombardment of energetic particles carried by magnetospheric currents. In this case, the putative hot spot  of PSR J0250$+$5854 should be colder than 200 eV.

An even more exotic class is composed by transient/intermittent radio sources that, when are active, show strongly polarized coherent radio pulses on timescales of thousands of seconds. The first and more famous object of this class is GLEAM-X J162759.5$-$523504.3, that showed pulsating radio emission with a period $P\approx18$ minutes \citep{2022Natur.601..526H}, then GPM J1839$-$10 ($P\approx 22$ minutes \citep{2023Natur.619..487H}), ASKAP J1935+2148 ($P\approx 54$ minutes \citep{2024NatAs...8.1159C}), \rev{GLEAM-X J0704$-$37 ($P\approx 2.9$ hours \citep{2024ApJ...976L..21H}) and ASKAP J1839$-$0756 ($P\approx 6.45$ hours \citep{2025arXiv250109133L})} followed. The period derivative of these  objects has not been measured yet, but the upper limits set incredibly strong $B\pdx{dip}<10^{15-16}$ G. The variable flux and pulse profile, and the huge radio luminosity $\gtrsim\!\edot_{\rm rot}$, indicate that the emission is not generated purely by spin down but that an additional source of energy, possibly magnetic, is needed.

The properties of these ultra-long period radio sources resemble in some sense those of the 6.7 hr pulsating source at the center of SNR RCW 103 \citep{2016ApJ...828L..13R,2016MNRAS.463.2394D}, and those of the radio active magnetars \citep{2022ApJ...934..184R}. They have also been considered as candidate fast radio bursts progenitors \citep{2020MNRAS.496.3390B}, especially after the discovery of repeating fast radio bursts (see, e.g., the case of FRB 180916.J0158+65 having a periodicity of about 16.35 days \citep{2020Natur.582..351C}).

The nature of ultra-long period radio sources is still debated. They could be magnetars that emit coherent radio emission despite extreme values of $P$ and $\pdot$ thanks to alternative sources of power, such as magnetospheric twists powered by plastic crustal motion \citep{2024MNRAS.533.2133C}, or they could be magnetic white dwarfs with dipolar spin-down emission enhanced by the intra-binary shock with the wind of the companion star \citep{2024ApJ...961..214R}, similarly to the models that explain the radio emission of AR~Sco ($P \approx 1.95$ min in a 3.5 hr orbit \citep{2016Natur.537..374M}) and J1912$-$4410 ($P \approx 5.3$ min in a 4 hr orbit \citep{2023NatAs...7..931P}). However, no optical/IR counterparts of the ultra-long period radio sources have been detected so far, ruling out the existence of a hot companion star \citep{2023MNRAS.520.1872B}. \rev{One remarkable exception is the recently-discovered GLEAM-X J0704$-$37: its high Galactic latitude and its M3-dwarf star companion, detected in the optical band, exclude the magnetar interpretation for this system \citep{2024ApJ...976L..21H}.}

\subsection{Links between the classes of \xd and RPP}\label{sec:link}
The most recent discoveries in the X-ray manifestations of INSs have narrowed the differences between the classes of \xd and RPP. These concern the radio high-B pulsar \psrj, that has similar timing and spectral properties of \xds, and the detection of non-thermal emission of two \xds.

\psrj is slowly rotating ($P \approx 3.4$ s), highly magnetized ($B\pdx{dip} = 3\times10^{13}$ G)
radio pulsar: its timing parameters are in the range of those of the \xds, but it does show radio pulsations. The similarity with the \xds was reinforced by X-ray observations with the \chandra \citep{2011ApJ...743..183S} and \xmm \citep{2019A&A...627A..69R} satellites, that revealed a purely soft thermal spectrum (temperatures $kT_1 = 74_{-11}^{+6}$ eV and $kT_2 = 140_{-20}^{+40}$ eV) plus an absorption Gaussian line with the line placed at $E=390_{-30}^{+20}$ eV and with a broadening of $\sigma=80_{-20}^{+30}$ eV. The inferred magnetic field is $B \approx 5\times 10^{13}$ G (see Eq.\ \ref{eq:Bcycp}), in good agreement with the dipole magnetic field.

\begin{figure*}[!ht]
\centering
\includegraphics[width=0.45\textwidth]{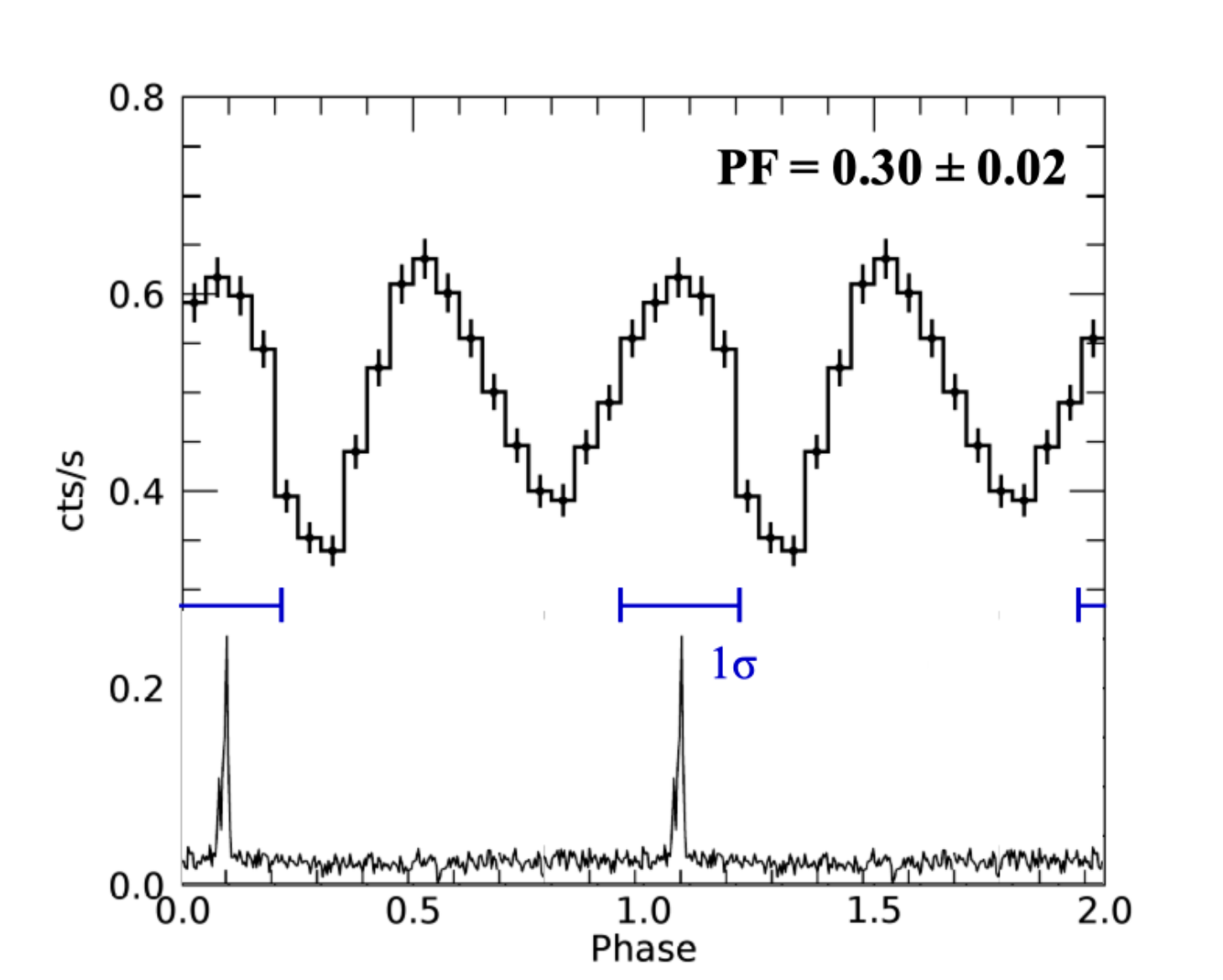} ~\qquad
\includegraphics[width=0.45\textwidth]{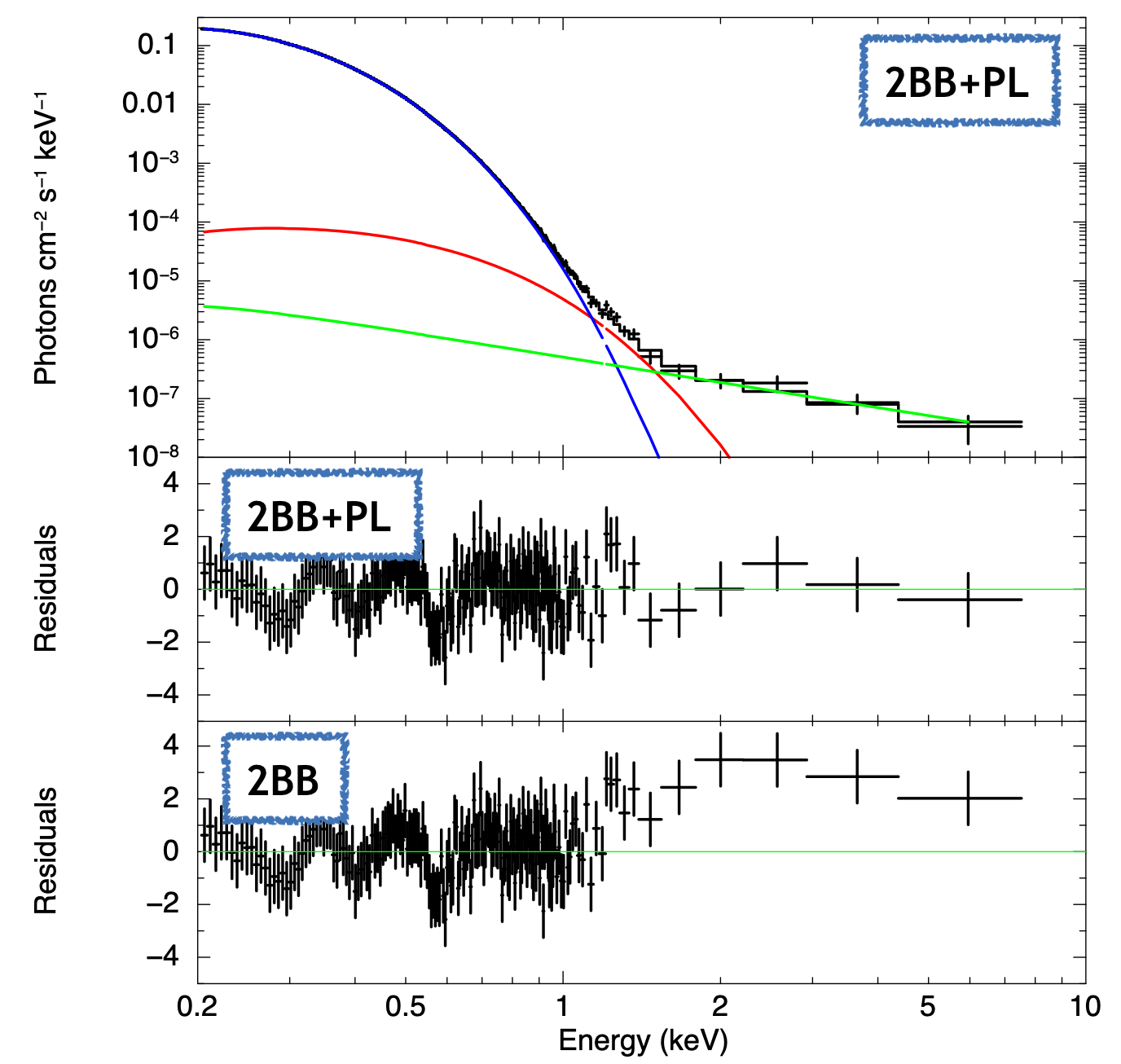}
\caption{\small Left panel: X-ray and radio pulse profile of \psrj. Adapted from Rigoselli et al.\ 2019 \citep{2019A&A...627A..69R}. Right panel: X-ray spectrum of \rxj obtained with \xmm data. The best-fitting model with two blackbody (blue and red) and a power-law (green) components is superimposed in the top panel. The two bottom panels show the residuals for the two-blackbody plus power law fit (2BB+PL) itself and for the two blackbody (2BB) fit, which yield unsatisfactory residuals at $E > 1$ keV. On the contrary, the residuals below $0.6$ keV are at a level $<\!2\%$, well below the instrument effective area uncertainties. Adapted from De Grandis et al.\ 2022 \citep{2022MNRAS.516.4932D}.} 
\label{fig:psrxd}
\end{figure*}

The X-ray pulse profile is sinusoidal and double-peaked, with a pulsed fraction of $30 \%$ (see Figure~\ref{fig:psrxd}, left panel). It cannot be easily reproduced by simple models based on blackbody emission, and the best match with the data was obtained assuming emission from two antipodal hot spots with an effective temperature of 0.5 MK reprocessed by a magnetized atmosphere model. The inferred geometry $(\Omega - \mu) \approx (\Omega - \mathrm{LOS})$ (where $\Omega$ and $\mu$ are the rotation and the magnetic axes, respectively, and the LOS is the direction of the line of sight) allows the detection of radio emission, while it is at variance with those of other XDINSs (RX J0720.4$-$3125 \citep{2011A&A...534A..74H} and RX J1308.6$+$2127 \citep{2017A&A...601A.108H}). This discrepancy might explain why \psrj is detected in the radio band, while the two \xds are not.

The soft X-ray pulsar \rxj  is the brightest ($F_X \approx 1.5 \times 10^{-11}$ \flux \citep{1996Natur.379..233W}) and closest ($d=123^{+11}_{-15}$ pc \citep{2010ApJ...724..669W}) member of the \xd class.
Its pulsation, with $P \approx 7.05$ s  \citep{2007ApJ...657L.101T} and $\pdot \approx 3\times10^{-14}$ \sps \citep{2008ApJ...673L.163V}, was detected despite a very small pulsed fraction $\rm{PF} \approx 1.2\%$. The timing parameters yield $\edot\pdx{rot}=3.4\times10^{30}$ \lum, $\tau_c = 3.7 \times 10^6$ yr and $B\pdx{dip} = 1.5\times10^{13}$ G.

The brightness, simple spectrum, and steadiness of its emission make \rxj an ideal target for the calibration of \xmm, that observed it about every six months since 2002. Using all the 2002--2022 data from the \xmm/EPIC-pn camera, De Grandis et al.\ 2022 \citep{2022MNRAS.516.4932D} obtained a $0.3-7.5$ keV spectrum having 1.43 Ms of net exposure time, that showed a hard excess with respect to the pure blackbody emission \citep{2017PASJ...69...50Y,2020ApJ...904...42D}. They report a best fit made up of the sum of two blackbodies (temperatures $kT_1 = 61.9 \pm 0.1$ eV and $kT_2 = 138 \pm 13$ eV, radii $R_1 = 4.92_{-0.06}^{+0.04}$ km and $R_2  = 31_{-16}^{+8}$ m) and a power law (photon index $\Gamma = 1.4_{-0.4}^{+0.5}$, flux in the $2-8$ keV band of $(2.5_{-0.6}^{+0.7})\times 10^{-15}$ \flux (see Figure~\ref{fig:psrxd}, right panel).

The luminosity of the non-thermal component corresponds to $10^{-3}$ times the spin-down power $\edot\pdx{rot}$. This value is consistent with what is observed in rotation powered X-ray pulsars with higher $\edot\pdx{rot}$ \citep{2002A&A...387..993P}. Hints for pulsations above 2 keV were also reported, so that a magnetospheric origin for this component appears as the most natural option. 

De Grandis et al.\ 2022 \citep{2022MNRAS.516.4932D} analyzed also 20 years of \xmm data of RX J0420.0$-$5022, the \xd having the second highest $\edot_{\rm rot}$ of the class.
They found a hard excess that can be fit either with a second blackbody of a power law, or their sum. Also in this case, the putative non-thermal component would have an efficiency $L\pdx{PL}/\edot\pdx{rot} \sim 10^{-3}$. 
If we assume the same efficiency, we expect that the other five \xds could show a similar non-thermal component with a flux $F\pdx{PL} \lesssim 10^{-16}$ \flux. This level of flux is beyond the sensitivity of the current facilities, but it maybe a target for future X-ray missions, such as \emph{NewAthena} 
\citep{2025arXiv250103100C}.

\subsection{Increasing the sample of INSs}\label{sec:tins}

The storage of several hundreds of ks of \xmm and \chandra pointings, as well as the increased sensitivity of the all sky survey in the soft X-ray band provided by \erosita, allowed us to discover new INS candidates even in the absence of pulsations. This could be caused by an unfavorable orientation of the rotation and magnetic axes, or by the intrinsic lack, or faintness, of non-thermal magnetospheric emission.

\begin{figure*}[ht]
\centering
\includegraphics[height=0.4\textwidth]{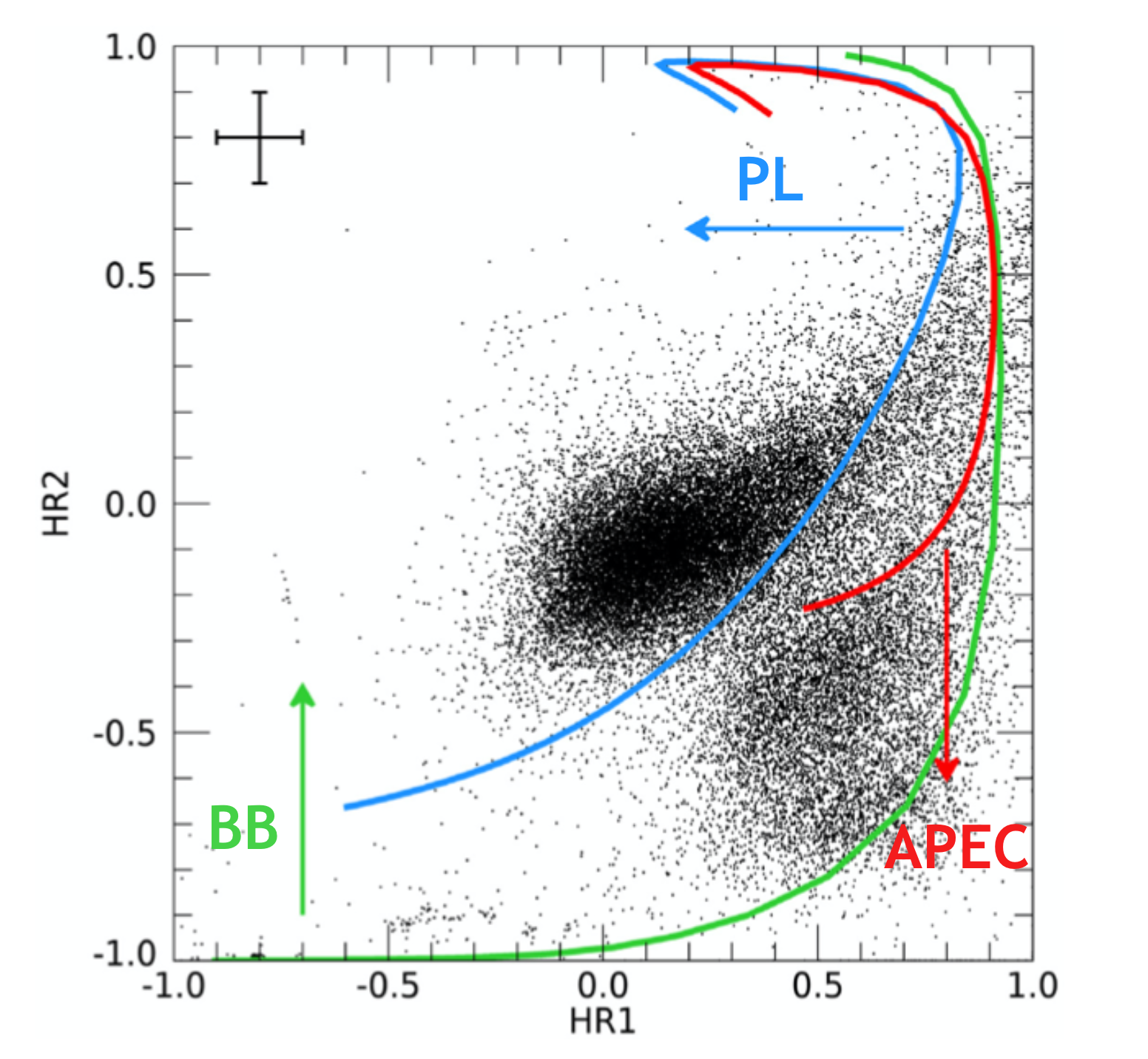}
\includegraphics[height=0.4\textwidth]{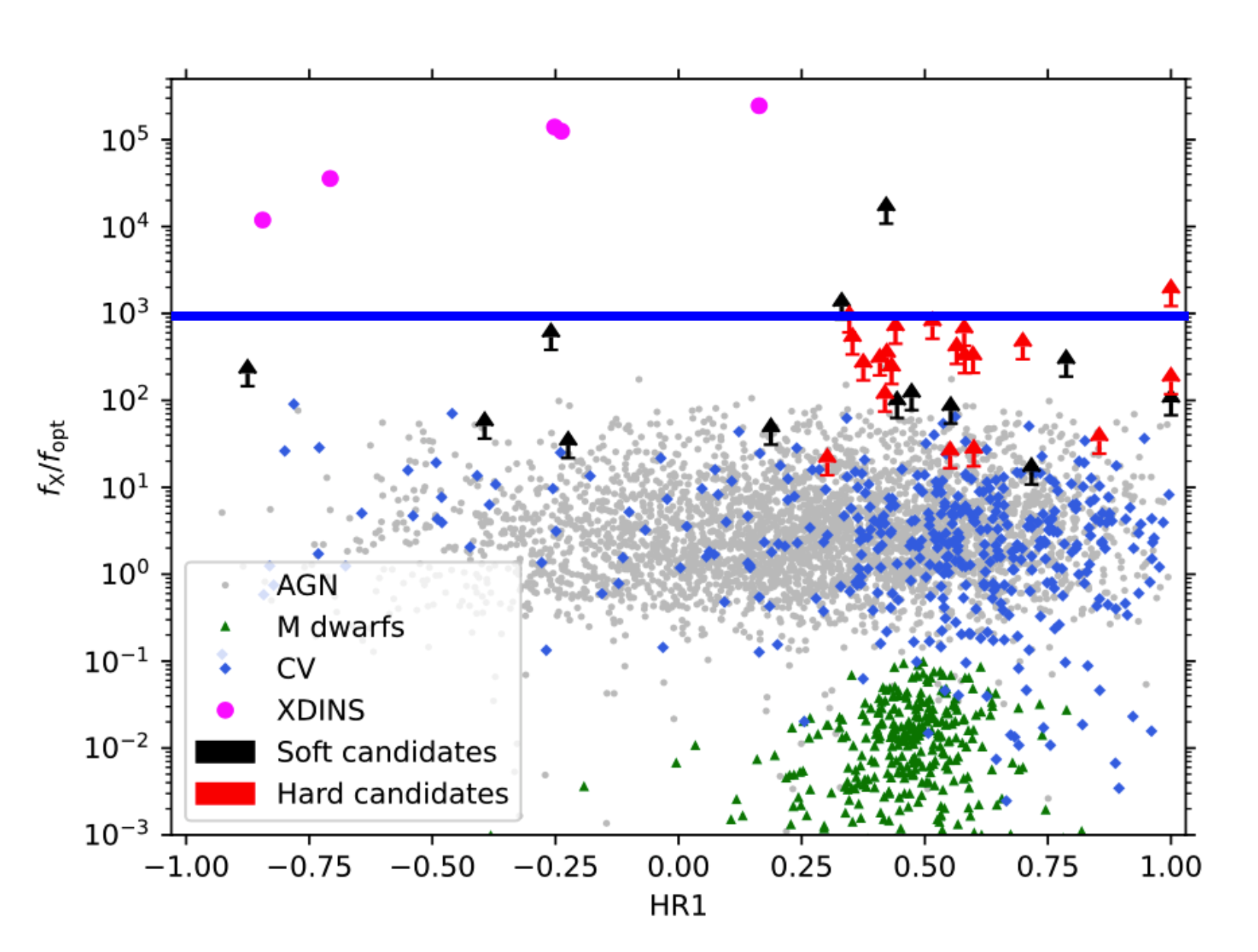}
\caption{\small Left: HR$_1-$HR$_2$ plot of the the point-like 4XMM-DR9 sources. The considered energy ranges are $0.2-0.5$, $0.5-1$ and $1-2$ keV. Colored lines represent the expected values for analytic models (blue line: \textsc{pl}, $\Gamma\lesssim4$; red line: \textsc{apec}, $kT\lesssim1.05$ keV, 0.5 solar abundance; green line: \textsc{bb}, $kT=0.05$ keV) convolved with the instrumental response function. All the lines are obtained varying \nh from $0$ to $10^{23}$ \col using the \textsc{tbabs} model. Credits to Rigoselli et al.\ 2022 \citep{2022MNRAS.509.1217R}.
Right: $F_X/F_O$ as a function of HR$_1$ for all the soft (black) and hard (red) candidates from the \erosita first catalog. The known \xds are also shown with magenta dots. The most prevalent contaminants are M~dwarfs (green), cataclysmic variable stars (blue) and AGNs (grey), that are all below a threshold value of $10^3$ (blue line). Credits to Kurpas et al.\ 2024 \citep{2024A&A...687A.251K}.}
\label{fig:TINS}
\end{figure*}

Among the several hundreds of thousands of unassociated X-ray sources, thermally-emitting INSs are point like and have a constant long-term emission, characterized by a soft, thermal spectrum, with a high X-ray-to-optical flux ratio ($F_X/F_O$). The soft thermal spectrum can be constrained even without a spectral fit using a hardness-ratio plot in the soft X-ray band (Figure~\ref{fig:TINS}, left panel).
This selection excludes the most abundant class of X-ray sources, i.e.\ the active galactic nuclei (AGN). The second class is composed by main sequence stars, bright in the X-rays but even brighter in the optical band. Therefore, a cut of $F_X/F_O>10^3$ can select only the INSs (Figure~\ref{fig:TINS}, right panel).

The most promising INS candidates are 
2XMM J104608.7$-$594306 \citep{2009A&A...498..233P,2015A&A...583A.117P} and
4XMM J022141.5$-$735632 \citep{2022MNRAS.509.1217R,2022A&A...666A.148P},
plus about 30 sources announced in the recent data release of the first \erosita catalog \citep{2024A&A...687A.251K}.
The ultimate proof for a neutron star association is the detection of pulsation: eRASSU J131716.9$-$402647 \citep{2024A&A...683A.164K} and eRASSU J065715.3+260428 \citep{2025arXiv250107347K} were targeted by \xmm and \nicer, and pulsations with $P\approx 12.76$ s and $\pdot<8\times10^{-11}$ \sps (J1317) and $P\approx 0.261$ s and $\pdot=6_{-4}^{+11} \times10^{-15}$ \sps (J0657) were detected. Absorption features in the thermal X-ray spectrum were measured at several hundreds of eV 
implying a magnetic field of $B\sim10^{13-14}$ G. Deep optical observations with VLT put even more constraining magnitude limits of $m>27.5$ (J1317) and $m>27.3$ (J0657), implying $F_X/F_O>10^4$.

Among the thermal-emitting INSs, Calvera (1RXS J141256.0$+$792204) is one of the most enigmatic sources because its timing and spectral properties do not fit easily with those of the known classes of INSs. 
It was discovered in the \rosat all sky survey as a soft X-ray source  with high $F_X/F_O$, qualifying it as an INS candidate \citep{2008ApJ...672.1137R}.   Its spectral properties (emission of thermal X-rays only, absence of radio and $\gamma$-ray counterparts) resemble those of \xds and CCOs, and for these reasons it was initially considered as a possible new member of the ``Magnificent Seven''; however, it was later discovered that Calvera has a spin period of 59 ms \citep{2011MNRAS.410.2428Z} and is spinning down at a rate $\pdot \approx 3.2\times10^{-15}$ \sps  \citep{2013ApJ...778..120H}. These timing parameters give a characteristic age $\tau_c = 2.9\times10^5$~yr and  a dipole magnetic field $B_{\rm dip}=4.4\times10^{11}$ G, similarly to middle-aged RPPs. For this reason it was nicknamed ``Calvera''. 

\begin{figure*}[ht]
\centering
\includegraphics[width=1\textwidth]{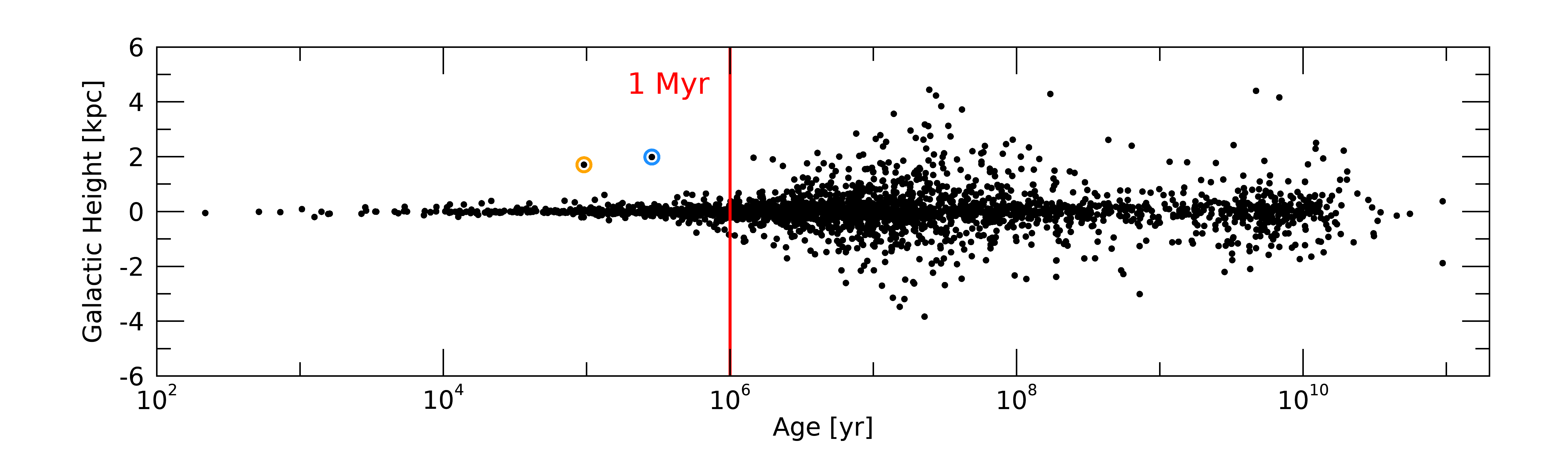}
\caption{\small Height on the Galactic plane as a function of characteristic age for all the Galactic INSs (excluded those in globular clusters). The majority of stars younger than 1 Myr is less than $\sim\!500$ pc distant from the plane; the only two exceptions are PSR J1951+4724 \rev{(orange, $b\approx10\deg$, $d\approx9.4$ kpc or $d\approx6.1$ kpc inferred from the radio dispersion measure and the models YMW16 \citep{yao17} and NE2001 \citep{cor02}, respectively)} and Calvera (blue, $b\approx37\deg$, $d\approx3.3$ kpc inferred from fitting the thermal X-ray spectrum \citep{2021ApJ...922..253M}). The data are taken from the ATNF Pulsar Catalogue, version 2.5.1 \citep{2005AJ....129.1993M}.}
\label{fig:calvera}
\end{figure*}

A magnetized hydrogen atmosphere model, covering the entire star surface and having an anisotropic temperature map, provides a good description of the phase-resolved spectra and energy-dependent pulsed fraction observed by \nicer\ \citep{2021ApJ...922..253M}. The inferred distance $d \approx 3.3$ kpc, coupled with a Galactic latitude $b \approx +37\deg$, provides an unusually high height above the Galactic disk ($z\approx 2$ kpc, see Figure~\ref{fig:calvera}). This supports the idea that Calvera was born in the Galactic halo, most likely from the explosion of a run-away massive star or, possibly, in a more unusual event involving a halo star, such as, e.g., the accretion induced collapse of a white dwarf.

Zane et al.\ 2011 \citep{2011MNRAS.410.2428Z} reported the presence of diffuse X-ray emission about $13^\prime$ west of Calvera, with spectral properties consistent with a SNR; recently, radio \citep{2022A&A...667A..71A} and $\gamma$-ray \citep{2022ApJ...941..194X,2023MNRAS.518.4132A} counterparts of this remnant were discovered. 
The association between the pulsar and the SNR was recently confirmed with the measurement of a proper motion of $78.5 \pm 2.9$ mas\,yr$^{-1}$ pointing away from the center of the ring \citep{2024ApJ...976..228R}.
These findings imply that Calvera is much younger than inferred from its timing parameters, reinforcing the similarities with the CCO class. 

\section{Multipolar and toroidal magnetic field}\label{sec:multi}
In this section I will report the growing evidence that the magnetic field has complicated field lines (multipolar and toroidal components) in the crust and above the surface also in ordinary RPPs: these concern the polar cap size (Section~\ref{sec:pc}), the presence of absorption features in the soft X-ray band (Section~\ref{sec:abs}), and the thermal surface map that can be inferred from timing and spectral analysis (Section~\ref{sec:thx}).

\subsection{Polar cap size}\label{sec:pc}

The study of neutron star heating by backward-accelerated particles is of particular interest for the study of particle acceleration in the pulsar magnetosphere \citep{2001ApJ...556..987H,2002ApJ...568..862H,1986ApJ...300..500C,1986ApJ...300..522C}.
In the classical dipolar scenario, the hot spots are identical and antipodal, are situated in correspondence of the magnetic poles and have a size $R\pdx{dip}= \sqrt{2\pi R_*^3/Pc} \approx 145 \,P\pdx{1\,s}^{-1/2}$~m.
Several \xmm and \chandra observations have been devoted to measure the characteristics of the hot spots of the brightest old pulsars, that have cooled down enough so that this contribution can be disentangled from the emission of the cooling surface, no longer relevant in the X-ray band. The emitting radii of the hot spots, inferred with blackbody fits, are a few meters only, and not all the pulse profiles are symmetric \citep{2008ApJ...686..497G,mis08,her13,her17,2018ApJ...865..116I}. 

It was proposed \citep{2008ApJ...686..497G,2003A&A...407..315G} that such a small emitting area can be explained if we consider that, close to  the star surface, the magnetic field is stronger than the dipolar field and described by multipolar components: 
\begin{equation}
    B\pdx{PC} = B\pdx{dip} \left( \frac{R\pdx{dip}}{R\pdx{PC}} \right)^2 \approx 10^2 B\pdx{dip}
    \label{eq:Bmulti}
\end{equation}
and the conservation of the magnetic flux through the polar cap area explain why we observe $R\pdx{PC} \sim 10-40$ m.
The asymmetry of the pulse profile also denotes that the geometry of the magnetic field close to the surface may be off-centered, and this would imply two caps that are non-identical, non-antipodal, and out of phase with respect to the radio main peak.
Recent results obtained with \nicer \citep{2019ApJ...887L..26B,2019ApJ...887L..23B,2019ApJ...887L..21R,2023A&A...680A..93P} also report evidence for non-dipolar magnetic fields, but they refer to the older millisecond pulsars that have $B\pdx{dip}\sim10^9$ G.

These discrepancies can also be addressed using more realistic thermal models, accounting for the effects of magnetic field on either a condensed surface or an atmosphere. It is well known that atmospheric models are \rev{spectrally harder} than blackbodies, thus yielding best-fit temperatures a factor two lower and, as a consequence, larger emitting areas. For three pulsars, the latter are consistent with the dipole polar caps (PSR B0950$+$08 \citep{zav04}, PSR B0823$+$26 \citep{her18}, PSR B0943$+$10 \citep{2019ApJ...872...15R}). 
However, these models depend on many more parameters than a simple blackbody, such as the chemical composition, the star mass and radius, the magnetic field, the system geometry, that in most cases are unknown and difficult to constrain due to relatively faint emission.

\subsection{Absorption lines}\label{sec:abs}

As said in Section~\ref{sec:cooling}, cyclotron features are detected in the (thermal) X-ray spectra of particular classes of INSs, that have the appropriate magnetic field to produce electron or proton absorption lines, respectively.
On the contrary, for the ordinary pulsars (with $B\pdx{dip}\sim10^{12}$~G) neither of the two conditions is met at the star surface and hence no analogous cyclotron features are expected in soft X-rays (Eqq.\ \ref{eq:Bcyce} and \ref{eq:Bcycp}).

Nevertheless, in the last few years some ordinary RPPs 
(PSR J1740$+$1000 \citep{2012Sci...337..946K}, PSR B1133$+$16 \citep{2018A&A...615A..73R}, and PSR B0656$+$14 \citep{2018ApJ...869...97A})
showed evidence for the presence of features in their X-ray spectra at about $\sim\!500$ eV.  The case of PSR J1740$+$1000 is extremely intriguing because, after a glitch occurred in 2012, the X-ray spectrum has changed and the absorption lines seem to have disappeared \citep{2022MNRAS.513.3113R}.

If the electrons are responsible for such features, they must be located high in the magnetosphere, at several stellar radii above the stellar surface where the dipole field is weaker ($B \propto (r/R_*)^{-3}$). Thus, the absorbed X-ray photons can be produced either on the surface or in the close magnetosphere. 
Alternatively, if the lines are attributed to protons close to the star surface, then a magnetic field $B = 4.5 \times 10^{13} \mathrm{~G} \approx 10 \times B\pdx{dip}$ is needed. 

\subsection{Surface temperature maps}\label{sec:thx}

In a strong magnetic field, electrons move more easily along the field lines than across them, making heat conductivity anisotropic. As a consequence, the magnetic field geometry determines the direction of heat conduction in INSs. In the radial magnetic field region of the envelope, heat is transported efficiently along the radial direction, establishing a thermal connection between the surface and the inner crust and core. Conversely, the regions with a nearly tangential magnetic field exhibit insulation and thermal disconnection from the hot core.

There are two quantities that are directly correlated to the temperature map and that can be inferred from X-ray observations: the pulsed fraction (PF) of the thermal component, and the deviation from a one-temperature thermal spectrum. 

Several works \citep{gre83,tur13,2013MNRAS.434.2362P} showed that poloidal-dominated configurations can only yield symmetric pulse profiles and very low PFs unless the local emission is highly beamed. On the other hand, the anisotropies in the surface temperature profile induced by the presence of a strong, large-scale toroidal field generally produce single-peaked pulsed profiles, with a PF that can reach about the 50\% even for purely isotropic local emission \citep{2013MNRAS.434.2362P,per06a,per06b,gep06}.

The majority of INS thermal spectra can be fitted with, at least, two blackbodies, that mimic the two extremes of a more complex temperature distribution. Rigoselli et al.\ 2022 \citep{2022MNRAS.513.3113R} showed that, with a few notable exceptions, the ratios of emitting radii ($R\pdx{hot}/R\pdx{cold}$) and temperatures ($T\pdx{hot}/T\pdx{cold}$) are very similar for all the families of INSs and are in the range $R\pdx{hot}/R\pdx{cold} \approx 0.03-0.3$ and $T\pdx{hot}/T\pdx{cold} \approx2$. Thus it is difficult to interpret the hotter and colder components in terms of emission from hot spots and from the whole surface ($R\lesssim200$ m and $R\sim12-16$ km, respectively), nor in terms of purely dipolar magnetic field, where a temperature contrast is expected to be below 1.5 (see also Yakovlev 2021 \citep{2021MNRAS.506.4593Y}).

\section*{Acknowledgments}
MR acknowledges INAF support through the Large Grant ``Magnetars'' (P.I. S. Mereghetti) of the ``Bando per il Finanziamento della Ricerca Fondamentale 2022''.

\providecommand{\href}[2]{#2}\begingroup\raggedright\endgroup

\vfill
\pagebreak


\end{document}